\documentstyle[prb,aps,epsfig]{revtex}

\begin{document}
\draft
\title{Surface Resistance Imaging with a Scanning Near-Field Microwave Microscope}
\author{D. E. Steinhauer, C. P. Vlahacos, S. K. Dutta, F. C. Wellstood, and Steven
M. Anlage$^{a)}$\footnotetext{$^{a)}$
Electronic mail: anlage@squid.umd.edu; color versions of the figures in this
paper are available at http://www.csr.umd.edu}}
\address{Center for Superconductivity Research, Department of Physics, University of\\
Maryland, College Park, MD 20742-4111 }
\maketitle

\begin{abstract}
We describe near-field imaging of sample sheet resistance via frequency
shifts in a resonant coaxial scanning microwave microscope. The frequency
shifts are related to local sample properties, such as surface resistance
and dielectric constant. We use a feedback circuit to track a given resonant
frequency, allowing measurements with a sensitivity to frequency shifts as
small as two parts in 10$^6$ for a 30 ms sampling time. The frequency shifts
can be converted to sheet resistance based on a simple model of the system.
\end{abstract}


There is a growing need to develop non-destructive microscopy techniques to
quantitatively measure the microwave properties of materials on a length
scale much less than the free space wavelength. For example, spatially
resolved measurements of complex conductivity would be of significant
utility for evaluating oxide superconducting and ferroelectric thin film
samples. Sensitivity to microwave and millimeter wave surface resistance and
dielectric constant have been previously demonstrated; for example, Bryant
and Gunn\cite{Bryant} used a coaxial resonator to measure semiconductor
resistivities on a 1 mm length scale. Waveguides\cite{Golosovsky} and
coaxial geometries\cite{Weiss}$^{-}$\cite{Vlahacos1} have also been used to
image conductivity and dielectric constant contrast. In this letter, we
describe the use of a microwave microscope with an open-ended coaxial probe
to quantitatively map the surface resistance of a metallic thin film.

\begin{figure}[htb]
\begin{center}
\leavevmode
\epsfxsize=8cm
\epsffile{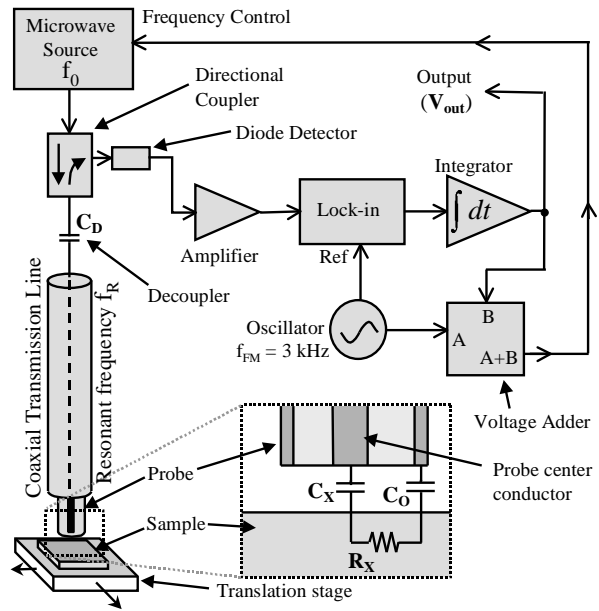}
\end{center}
\caption{Schematic of the near-field scanning microwave microscope. The
inset shows the probe tip and sample, with the probe-sample capacitance C$_X$
and the sample resistance R$_X$. The feedback circuit keeps the microwave
source locked onto a resonant frequency of the circuit consisting of a
coaxial transmission line and open-ended coaxial probe.}
\label{fig1}
\end{figure}

The key element in our system\cite{Vlahacos1}$^{,}$\cite{Anlage} is a 2
meter-long resonant coaxial transmission line (see Fig. \ref{fig1}). One end
of the line is connected to an open-ended coaxial probe and the other end is
weakly coupled to a microwave source via a capacitor $C_D$. Near-field
microwave energy at the exposed tip of the probe center conductor is coupled
to the sample. As the sample is scanned beneath the probe tip, the resonant
frequencies and quality factor Q of the open transmission line shift
depending on the surface properties of the region of the sample closest to
the probe's center conductor.\cite{Vlahacos1}$^{,}$\cite{Anlage} We measure
the microwave power reflected back up the transmission line with a diode
detector.\cite{HP diode} By using a fixed frequency source near one of the
resonances $f_R$, and scanning a sample underneath the probe, one can map
the reflected power and generate an image.\cite{Vlahacos1}$^{,}$\cite{Anlage}
However, this results in a convolution of two distinct contrast mechanisms:
the frequency shift of the standing wave resonances and the change in Q.

To disentangle these effects, we have developed a frequency-following
feedback circuit, as shown in Fig. \ref{fig1}. We frequency modulate the
source at a rate $f_{FM}\approx $ 3 kHz with a deviation of about $\pm $3
MHz and use a feedback loop to keep the average microwave source frequency
locked to a specific resonant frequency $f_R(t)$ of the microscope. To
accomplish this, the diode detector output voltage is amplified and sent to
a lock-in amplifier referenced at the frequency $f_{FM} $. The lock-in
output is time-integrated; this voltage signal $V_{out} $ is added to the $%
f_{FM}$ oscillator signal and fed to the frequency-control input of the
microwave source, (see Fig. \ref{fig1}) closing the feedback loop. With the
loop locked, the signal $V_{out}$ is proportional to the frequency shift. We
use a computer to record $V_{out}$, while controlling the sample position
using a two-axis translation stage. This data can be compiled into a
gray-scale frequency shift image.

One advantage to this circuit is its speed; the frequency shift can be
recorded while the sample is in motion beneath the probe. We typically
sample at a rate of 30 Hz, so that a $1\times 1$ cm sample can be scanned
with a resolution of 100 $\mu $m in about 5 minutes.

As shown in the inset to Fig. \ref{fig1}, we model\cite{Vlahacos1} the
interaction between the probe and sample as a capacitance $C_X$ between the
center conductor of the probe and the sample, and a resistance $R_X$
connected to the outer conductor via a second capacitor $C_O$. The
capacitance $C_X$ is determined by the area of the probe center conductor
and the height of the probe above the sample. Because of the large area of
the outer conductor, we take $C_O\gg C_X$; in this approximation, $R_X$ is
connected directly to the grounded outer conductor. For simplicity, we use a
parallel plate approximation for $C_X$, and assume that $R_X$ is equivalent
to the microwave surface resistance of the sample\cite{Klein}$^{,}$\cite
{Booth}. To model the system we also include the decoupler capacitance ($%
C_D\approx $ 0.17 pF) and attenuation in the transmission line.\cite{Gore}
We calculate shifts in the resonant frequencies as a function of the above
parameters using standard microwave circuit theory.\cite{Van Duzer}

We tested the system by measuring frequency shift as a function of sheet
resistance $R_X$. To vary $R_X$, we used a variable-thickness thin-film
oxidized aluminum sample. The aluminum film was deposited on a glass
substrate by thermal evaporation. The glass and the source were arranged so
that the source was much closer to one side of the substrate than the other,
producing a smooth variation in the thickness of the film. After oxidation
in air, the thickness of the conductive aluminum is less than 6 nm, and much
less than the skin depth at microwave frequencies, so that there is
essentially no reactive impedance presented by the sample.\cite{Klein} Using
two-point resistance measurements of constant-thickness strips of the
sample, we measured $R_X$ values which ranged smoothly from 58 $\Omega /\Box 
$ to 42 k$\Omega /\Box $ across the sample.

\begin{figure}[htb]
\begin{center}
\leavevmode
\epsfxsize=8cm
\epsffile{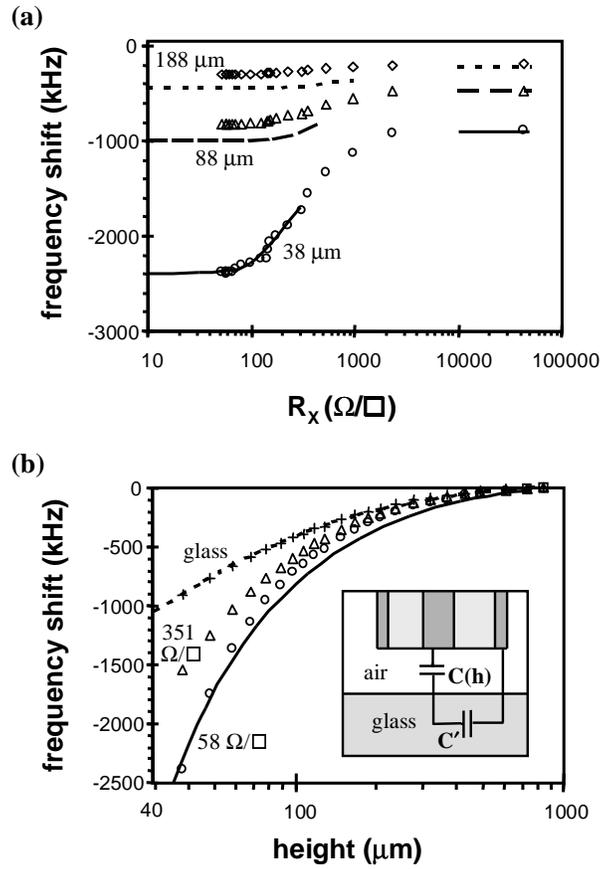}
\end{center}
\caption{The symbols show the measured frequency shift vs. (a) surface
resistance R$_X$, and (b) probe-sample separation. The lines indicate
expected frequency shift based on microwave circuit analysis of the system.
A probe with a 480 $\mu $m diameter center conductor was used, at a
frequency of 7.8 GHz. The inset is a diagram of our model for the
interaction between the probe and the sample when no film is present on top
of the glass substrate.}
\label{fig2}
\end{figure}

To measure the frequency shift, a probe with a 480 $\mu $m diameter center
conductor was positioned at various heights above the sample, and a
microscope resonance at 7.8 GHz was chosen. The data points in Fig. \ref
{fig2}a indicate measurements at different heights above the sample. For
decreasing $R_X$ at a fixed height, the resonant frequencies shift downward.
This can be understood as follows: as $R_X$ decreases, the boundary
condition of the resonator tends from an open circuit toward a short
circuit, changing the circuit from a half-wave resonator toward a
quarter-wave resonator, and lowering the resonant frequency. We note that
the graph of frequency shift vs. $R_X$ flattens out for low $R_X$,
indicating reduced sensitivity to sheet resistance in this region. In
addition, the curves for smaller heights have steeper slopes, demonstrating
that the sensitivity to $R_X$ increases as the probe-sample separation is
reduced.

For comparison, the expected curves from a microwave analysis of the circuit
are shown as lines in Fig. \ref{fig2}a. For $R_X<377\ \Omega /\sqrt{\epsilon
_r}$, where $\epsilon _r$ is the dielectric constant of the glass substrate,
the aluminum conductance dominates; in this region, we ignored the presence
of the substrate.\cite{Klein} Because of the large diameter of the probe,
there is an uncertainty in the height of the center conductor relative to
the sample. Accordingly, we fitted the height for one data set (the 38 $\mu $%
m set) and then used the measured heights relative to this set for the other
data sets. For low $R_X$, the 38 $\mu $m fit agrees quite well with our
experimental data, while the curves at 88 and 188 $\mu $m don't agree as
well; this discrepancy may be due to a breakdown of the parallel plate
approximation as the sample-probe separation approaches the radius of the
probe center conductor.

We also tested the frequency shift as a function of height for various
values of $R_X$ (see Fig. \ref{fig2}b). Notice that as the probe is moved
closer to the sample, the resonant frequencies drop. This is expected since
the increase in coupling capacitance between the probe and sample
effectively increases the length of the resonant circuit, which decreases
the resonant frequencies.\cite{Vlahacos1} As the lines in Fig. \ref{fig2}b
show, the model agrees with our experimental data for low values of $R_X$
where the effect of the glass substrate can be safely ignored, and for
heights small compared to the radius of the probe center conductor, where
the parallel plate approximation is valid.

Our simple model doesn't take into account the presence of the glass
substrate (a dielectric). For high $R_X$, if the glass were not present, the
resonant frequency shifts would tend to zero (defined as the resonant
frequency when the probe is $>$ 1 mm from the sample). This is not the case,
as shown by the experimental data in Fig. \ref{fig2}a. As shown in the inset
to Fig. \ref{fig2}b, we empirically model the effect of the glass by taking $%
R_X=0$, and modeling $C_X$ as two capacitors in series, $\frac 1{C_X}=\frac 1%
{C^{\prime }}+\frac 1{C(h)}$, with an air gap capacitance $C(h)=\frac{%
\epsilon _0A^{\prime }}h$. The two fitting parameters are $A^{\prime }$, an
effective area for the probe, and the fringe capacitance $C^{\prime }$
through the substrate. For the glass substrate the best fit to our model
gave $A^{\prime }=0.65A$ and $C^{\prime }=0.044$ pF, where $A$ is the area
of the probe's center. We used this modified model for plotting frequency
shifts for high $R_X$ in Fig. \ref{fig2}a, and for plotting the model curve
for glass in Fig. \ref{fig2}b. In principle, the empirical values of $%
A^{\prime }$ and $C^{\prime }$ could be used to determine the dielectric
constant of a material.

\begin{figure}[htb]
\begin{center}
\leavevmode
\epsfxsize=8cm
\epsffile{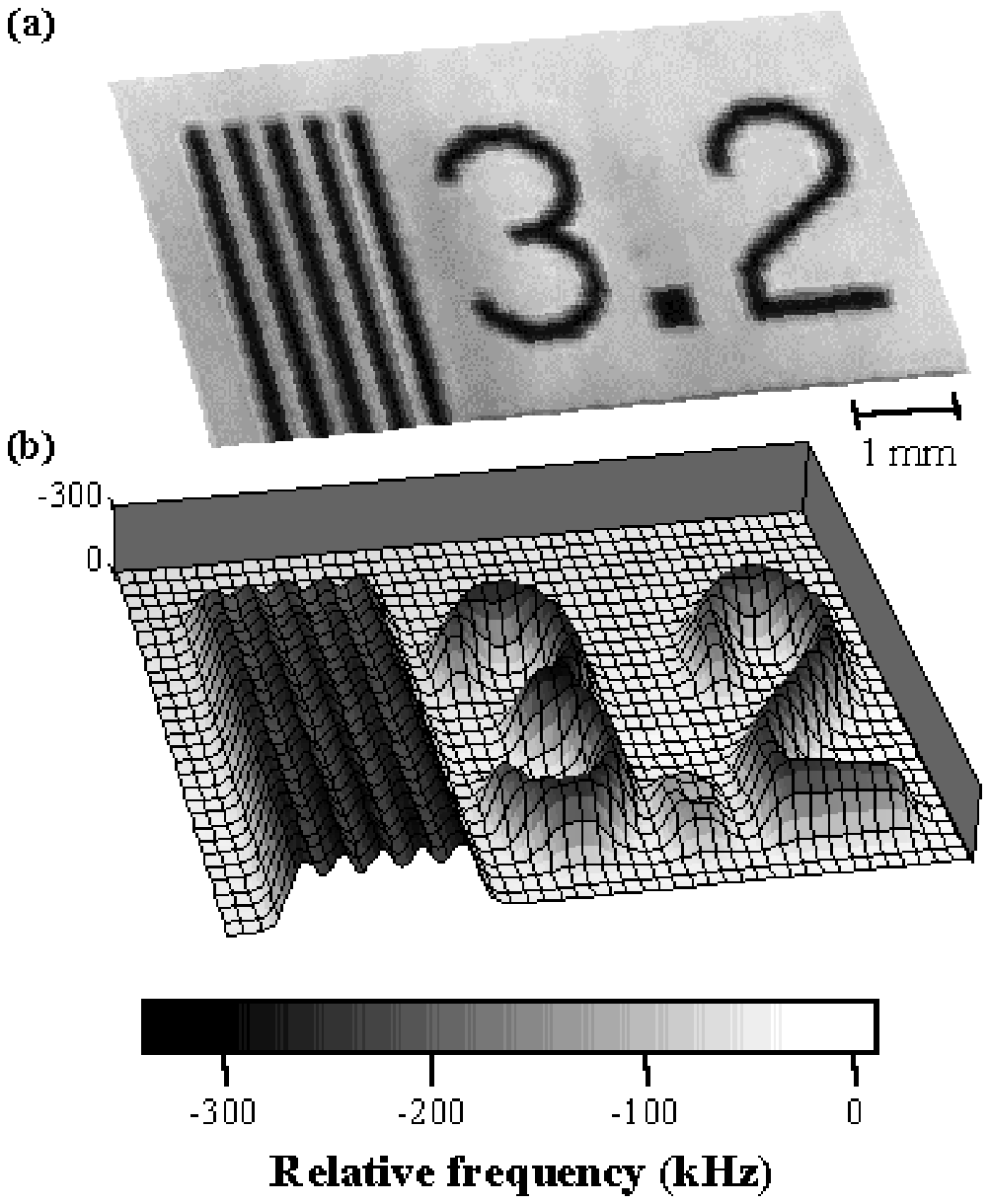}
\end{center}
\caption{(a) Optical photograph of a 3.2 lines/mm resolution target. The
target consists of a pattern of chromium thin film lines on glass. (b)
Frequency shift surface plot of the resolution target. A probe with a 200 $%
\mu $m diameter center conductor was used at a height of 5 $\mu $m and a
frequency of 10.7 GHz. The frequency shifts are relative to the resonant
frequency when the probe is over glass.}
\label{fig3}
\end{figure}

To demonstrate the imaging capabilities of the system, we imaged a 3.2
lines/mm NIST standard resolution target consisting of a patterned chromium
thin film on glass (see Fig. \ref{fig3}a). Figure \ref{fig3}b shows a
frequency shift image of the resolution target. As shown in the figure, the
resonant frequencies shifted downward by about 300 kHz when the probe moved
from over glass to over a chromium line. This is less than we would expect
for a uniform film with a sheet resistance of 10$\ \Omega /\Box $,
suggesting that the small width of the chromium lines (156 $\mu $m) relative
to the probe outer conductor diameter (860 $\mu $m) reduces the magnitude of
the observed frequency shifts from that expected for a uniform film. We note
that the 300 $\mu $m-wide decimal point in ``3.2'' is visible, even though
it is smaller than the probe outer conductor diameter; however, the decimal
point appears dimmer than the chromium lines and is doughnut-shaped, also
suggesting that the film pattern geometry has an additional effect on the
magnitude of the frequency shifts. Because of the concentration of the
fields near the inner conductor tip, the spatial resolution is determined by
the diameter of the inner conductor.\cite{Vlahacos1}

By observing circuit output noise, we find that with a sampling time of 30
ms the frequency-following circuit is sensitive to frequency shifts as small
as two parts in 10$^6$. Given the sensitivity to frequency shift and the
data in Fig. \ref{fig2}a, we can estimate the sensitivity to changes in $R_X$%
. For our 480 $\mu $m center conductor probe at a height of 38 $\mu $m, we
find $\Delta R_X/R_X=5\times 10^{-2}$ for $R_X=100\ \Omega /\Box $. To
achieve finer spatial resolution and maintain the same frequency shift
sensitivity to $R_X$, one could decrease the probe inner conductor diameter,
and decrease the probe-sample separation ($h$), in order to maintain the
same probe-sample capacitance. One limitation of this technique is that changes in the
probe-sample separation ($\Delta h$) must remain small relative to $h$
throughout the scan. If the sample is not planar, changes in probe-sample
capacitance will dominate at small $h$, thus setting a lower limit to how small $h$
can be for quantitative surface resistance imaging. A solution to this problem
would be an active
feedback system to keep the probe-sample separation constant.\cite{Rugar}

In conclusion, we have demonstrated the use of a frequency-following circuit
for microwave imaging of the surface resistance of metallic samples. The
sensitivity to surface resistance depends primarily on the probe-sample
capacitance. Our current system has the best surface resistance sensitivity
in the range from about 20 to 200 $\Omega /\Box $. In the low surface
resistance range, $R_X\lesssim 150\ \Omega /\Box $, our results on a uniform
sample show reasonable agreement with a simple model based on standard
microwave circuit theory.

This work has been supported by NSF-MRSEC grant \# DMR-9632521, NSF grant \#
ECS-9632811, one of us (S.M.A.) by NSF NYI grant \# DMR-9258183, and by the
Center for Superconductivity Research.

\newpage

\newpage

\newpage

\begin{references}
\bibitem{Bryant}  C. A. Bryant and J. B. Gunn, Rev. Sci. Instr. {\bf 36},
1614 (1965).

\bibitem{Golosovsky}  M. Golosovsky, A. Galkin and D. Davidov, IEEE Trans.
Microwave Theor. Tech., {\bf 44} (7), 1390 (1996).

\bibitem{Weiss}  S. J. Stranick, L. A. Bumm, M. M. Kamna and P. S. Weiss, in 
{\it Photons and Local Probes}, eds. O. Marti and R. M\"{o}ller, (Kluwer,
Netherlands, 1995), p. 221.

\bibitem{Xiang}  T. Wei and X.-D. Xiang, Appl. Phys. Lett. {\bf 68} (24),
3506 (1996).

\bibitem{Keilmann}  F. Keilmann, D. W. van der Weide, T. Eickelkamp, R. Merz
and D. St\"{o}ckle, Opt. Commun., {\bf 129}, 15 (1996).

\bibitem{Knoll}  B. Knoll, F. Keilmann, A. Kramer and R. Guckenberger, Appl.
Phys. Lett. {\bf 70} (20), 2667 (1997).

\bibitem{Vlahacos1}  C. P. Vlahacos, R. C. Black, S. M. Anlage, A. Amar and
F. C. Wellstood, Appl. Phys. Lett. {\bf 69} (21), 3274 (1996), and
references therein.

\bibitem{Anlage}  S. M. Anlage, C. P. Vlahacos, Sudeep Dutta and F. C.
Wellstood, IEEE Trans. Appl. Supercond. {\bf 7} (3), 3686 (1997).

\bibitem{HP diode}  Hewlett-Packard Company, Santa Clara, California:
microwave diode detector, part \# 8473C.

\bibitem{Klein}  N. Klein, H. Chaloupka, G. M\"{u}ller, S. Orbach, H. Piel,
B. Roas, L. Schultz, U. Klein and M. Peiniger, J. Appl. Phys. {\bf 67} (11),
6940 (1990).

\bibitem{Booth}  J. C. Booth, Dong Ho Wu and Steven M. Anlage, Rev. Sci.
Instrum. {\bf 65} (6), 2082 (1994).

\bibitem{Gore}  Private communication with W. L. Gore and Assoc., Inc.: for
ReadyFlex cable, attenuation constant $\alpha =$ 0.333 dB/ft at 7.8 GHz;
dielectric constant $\epsilon _r=$ 1.15; characteristic impedence $Z_0=$ 50 $%
\Omega $.

\bibitem{Van Duzer}  S. Ramo, J. R. Winnery and T. Van Duzer, {\it Fields
and Waves in Communication Electronics}, 3rd. ed. (Wiley, New York, 1994),
Chap. 5.

\bibitem{Rugar}  D. Rugar, H. J. Mamin and P. Guethner, Appl. Phys. Lett.
{\bf 55} (25), 2588 (1989).
\end{references}
\end{document}